\documentclass[preprint,iop,10pt,twocolumn] {revtex4-1}
\usepackage{graphicx}% Include figure files
\usepackage{dcolumn}% Align table columns on decimal point
\usepackage{amsmath}
\usepackage{bm}% bold math
\usepackage{natbib}

\usepackage[mathlines]{lineno}%
\begin{document}
%\preprint{APS/123-QED}
\title [Development of an interactive....] {Development of an interactive code for quick data analyses between STOR-M tokamak experimental plasma discharges.}
\author{Masaru Nakajima$^{a,\SS}$, Debjyoti Basu$^{a,b,\S}$, A.V. Melnikov$^{c,d}$, David McColl$^a$, Chijin Xiao$^a$}
\address{$^{a}$Plasma Physics Laboratory, University of Saskatchewan, Saskatoon, Canada}
\address{$^{b}$ Present address: University of Maryland, Baltimore County, MD-21250, USA}
\address{$^{c}$NRC ‘Kurchatov Institute’, 123182, Moscow, Russia.}
\address{$^{d}$ National Research Nuclear University MEPhI, 115409, Moscow, Russia.}
%\address{$^{c}$National Research Nuclear University MEPhI, 115409, Moscow, Russia.}
\email{$^{\S}$debjyotibasu.basu@gmail.com}
\email{$^{\SS}$masarun@usc.edu}
%\email{$^{\ddag}$melnikov\_07@yahoo.com}
%Second institution and/or address%Authors' institution and/or address%\\This line break forced with \textbackslash\textbackslash
%%}%

\graphicspath{{figs/}}

\date{\today}% It is always \today, today,
             %  but any date may be explicitly specified

\begin{abstract}
Saskatchewan Torus-Modified (STOR-M) is a small tokamak, well known for various fusion related basic experimental studies such as edge turbulent heating, different instabilities, AC (alternating current) tokamak operation, Ohmic H-mode triggering by the electrode biasing, fueling and momentum injection by Compact Torus(CT) injection, and effects of Resonance Magnetic Perturbations (RMP), among others. Some of those experiments require real time visualization of magnetic surface reconstructions either through EFIT or quick analyses and visualization of experimental data during experiments. Recently experimental studies of Geodesic Acoustic Mode(GAM) and zonal flows had been performed in STOR-M tokamak. The GAM experiments strongly require collection of fluctuations data from different Langmuir probes installed at different poloidal locations, but on the same magnetic surfaces. This is need of the adjustment of radial locations between discharges. It is therefore important to analyze and visualize the features of all probe data quickly during discharges. For this purpose, a Python code has been developed and used for  quick analyze of data. This article will describe the development of the code using Python and its use in detail.
\end{abstract}
%Valid PACS numbers may be entered using the \verb+\pacs{#1}+ command.

\pacs{Valid PACS appear here}% PACS, the Physics and Astronomy
%%                             % Classification Scheme.
%\keywords{Suggested keywords}%Use showkeys class option if keyword
                              %display desired
\maketitle
%\twocolumn
\section{Introduction}
Discovery of the concept of zonal flows has great scientific value in the field of basic turbulent plasma physics and magnetically confined plasma systems aiming to nuclear fusion in a controlled manner. A basic requirement to achieve self-sustaining burning plasmas in a magnetically confined fusion reactor is to control anomalous transportation of particle and energy. It is believed that anomalous transport happens due to small scale drift turbulent fluctuations in a magnetically confined toroidal system like tokamaks or stellarators. The anomalous transport can be controlled by radially sheared poloidal flow which is generated by either radially sheared mean or oscillating electric field. The poloidal mean flow may be generated by external means but in general the oscillating poloidal flows, confined within a small radial width, is a self-organized process which is known as 'zonal flows'. Theoretically and experimentally it has been shown that the drift wave turbulence is a sole cause of the origination of zonal flows. Two types of zonal flows have been found in a magnetically confined toroidal plasma system which are  i)near zero or very low frequency Zonal Flows(ZF) and  ii)Geodesic Acoustic Mode(GAM) which has been found only in toroidal geometry. The conventional GAM is featured by i) radially localized, ii) toroidally axisymmetric for potential and density fluctuations $(n=0;m=0)$ and $k_{\phi}=0$, iii) poloidally symmetric potential fluctuations, so, $(n=0)$, $k_{\theta}=0$ for potential fluctuations but asymmetric density fluctuation $(m=1)$ and $k_{\theta}=\pi$ for density fluctuations \cite{kn:Diamond,kn:Fujisawa}.
\par Since, the GAM mode is originated in a particular radius and normally confined within a narrow radial width, the experimental detection of GAM mode demands that diagnostics should collect data from the same magnetic surfaces where the mode is confined. The main issue of the tokamaks which are not facilitated by good feedback systems is the reproducibility of particular shot in a same magnetic surface with same characterizations. Similarly, the radial positions of GAM formation of a particular Tokamak plasma discharge may change shot by shot. In that case, the detection of radial locations of GAM mode should be done during real time operations through quick visualization of experimental data of the density and potential fluctuations whether they have been collected from the same magnetic surface by the help of highly advanced software like Equilibrium Fit(EFIT)\cite{kn:lao}. A high performance Python code has been developed for quick analysis the data taken by diagnostic system and derive mode structure between consecutive plasma discharges in STOR-M due to absence of EFIT.
\par In this article, detailed description of code development, preliminary experimental data analysis through the code during STOR-M discharges and brief description of mode detection will be discussed.

\section{Experimental Setup}
The STOR-M with circular plasma cross-section has major and minor radii 46cm and 12cm respectively. The plasma discharge duration is around 30 ms. The plasma boundary is defined by an elliptical stainless steel poloidal limiter, placed in a way that its internal major $(13 cm)$ and minor $(12 cm)$ radii have aligned along horizontally and vertically. Three Langmuir probe systems(top, radial-2, bottom) were installed through the top, equatorial and bottom ports of the vessel in the same poloidal cross-section. A Fourth one(radial-1) is placed equatorially, offset toroidally by $90^{0}$ in clockwise direction (top view) with respect to the other three sets. Detailed description of diagnostics arrangement can be found from the article of `GAM detection in STOR-M tokamak'\cite{kn:DB}. Now, the necessary and very important requirement is to determine that all Langmuir probes will collect data from the same magnetic field surface for the GAM detection experimental studies. The main purpose of this code development is to quick determination through fast analysis whether those Langmuir probes data are collected from the same magnetic surface. The detailed descriptions of the code development are given in following sections.
\section{Code Development}This code utilize the Graphical User Interface(GUI) feature in Python language, specifically for the STOR-M Tokamak experiments where probes data are recorded and stored in National Instruments(NI) digitize modules plugged in to a `PXI' crate and directly to the PSI slots in computers. There are around 120 channels in total. LabView software is used to manage all hardware and to transfer onboard data in digitizers to computers and save them in to the data files in the hard drives of computers after each shot. In experimental settings, the duty cycle time period between two discharges can be set to 4-10 minutes depending on particular experiments whereas the time to transfer and save data complete within a few seconds. It is imperative to quickly visualize and analyze the experimental data after every recorded STOR-M discharge for the complex experiment like GAM detection to determine whether satisfactory data were obtained, or the configurations of the diagnostics set up needs to be modified. The present development of code with interactive data analysis tools meet such requirements. In this section the logical frame of the application will be discussed. The resultant GUI window is presented by figure 1. The work flow chart of this application is shown in figure 2(a).
\begin{figure}[h]
\center
\includegraphics[width=260pt,height=150pt]{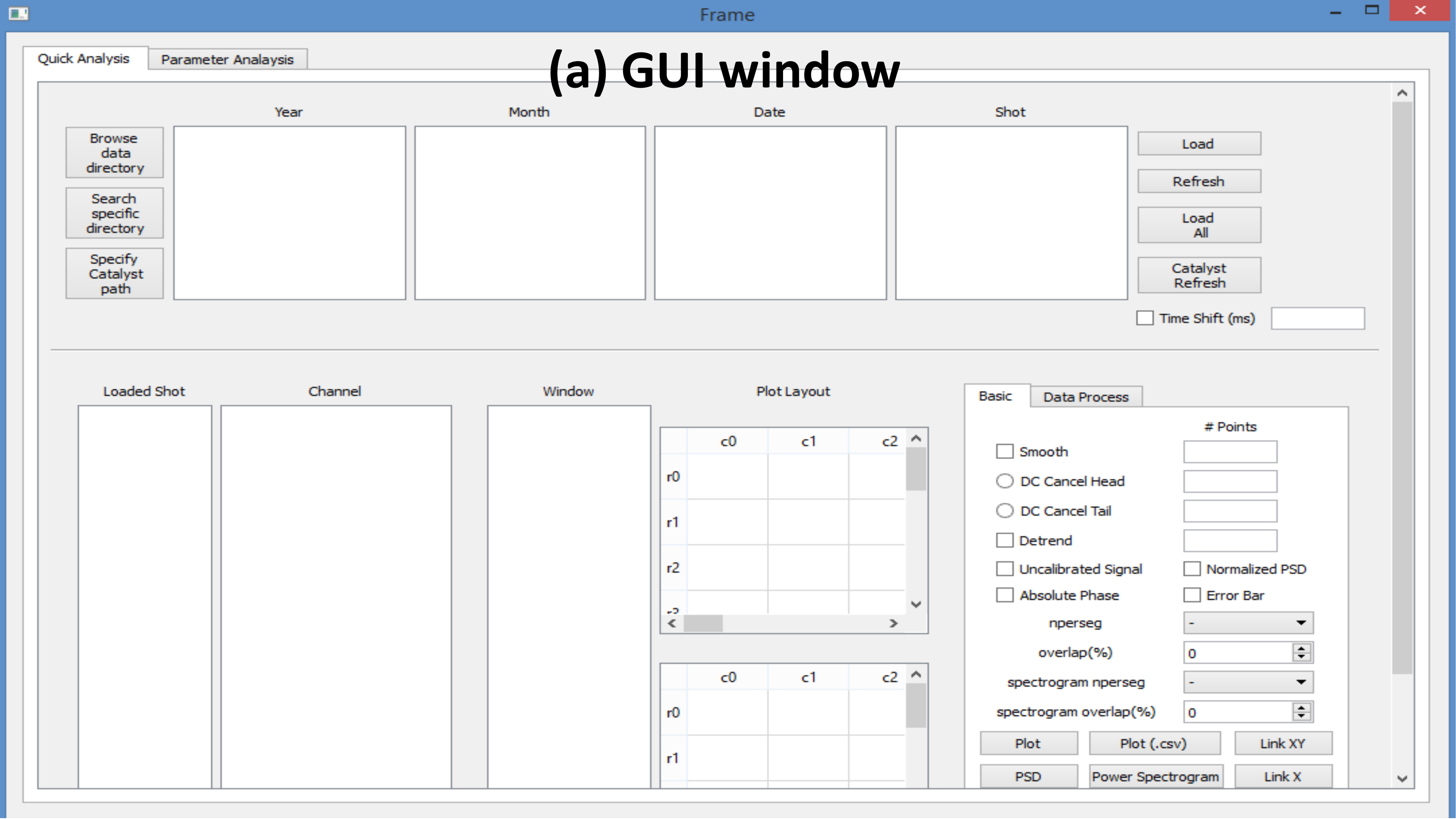}
\caption{It presents the image of GUI window of the code.} \label{fig:1}
\end{figure}

\begin{figure}[h]
\center
\includegraphics[width=260pt,height=150pt]{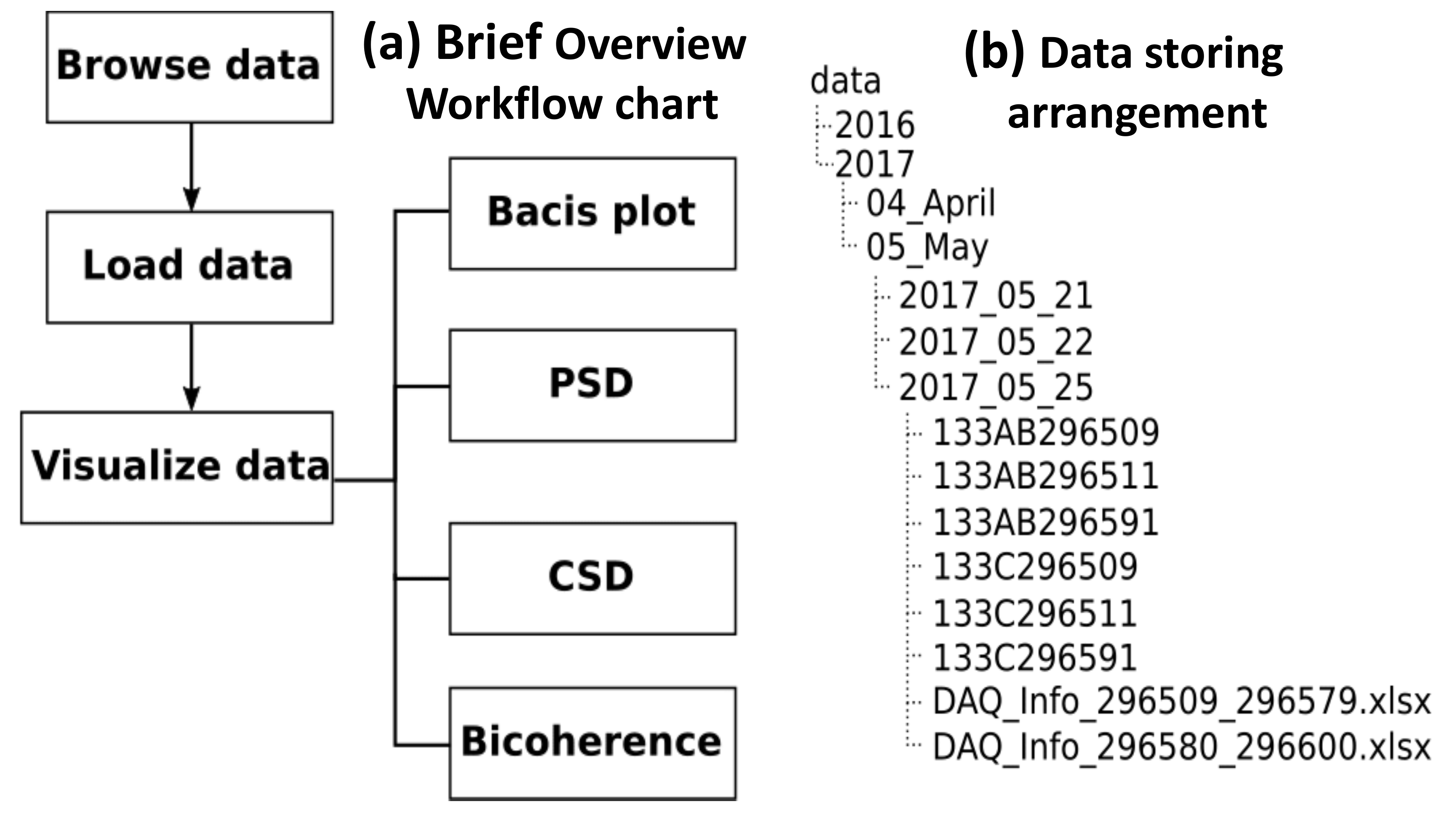}
\caption{It presents(a) work flow chart, (b)sequential organization of data store.} \label{fig:2}
\end{figure}

It describes how the programs run when the large amount of diagnostics data related to the experiment are store periodically after each STOR-M discharge in the pre-specified file in directories. So, the each step of the operations and their inter connections from $Browse Data\Rightarrow Load Data \Rightarrow Visualize Data$, the next level of analytical operations, including basic plots, `Power Spectral Density', `Cross Spectral Density', and `Bi-coherence'  are controlled by dedicated background programs which were solely developed. Others available applications which were used in a logical manner are graphical interface through the $\texttt{PyQt}$ package, a wrapper package for a GUI tool kit $\texttt{Qt}$~\cite{kn:pyqt},the plotting and visualization of data through $\texttt{PyQtGraph}$~\cite{kn:pyqtgraph}, a python toolkit for interactive
data visualization and the package $\texttt{SciPy}$ for spectral analysis. Figure 2(b) represent the sequence of data arrangement consecutively.
\par Each STOR-M plasma discharge is labeled by a shot number as a common experimental procedure for data storage. The data files for National Instrument digitizers are stored in the text files format with unique number synchronized with shot number corresponding to each set of digitizers within it. It is very important to develop a logical working procedure in the code which includes its methods of reading, interacting and operating on the stored experimental data. So, it requires a guiding soft file which can lead the code for its working flows. Here, the Excel book is the pilot file from which the code collects the instructions for loading data files. Each Excel pilot files were named in such a way that code can understand the range of shots number stored on the particular folder. The pilot files contain the descriptions for each measurement channel names, conversion factors, and the final unit of the measurement. Since multiple shots share a same measurement setup, the file names of the DAQ information files indicate the range of shot numbers for which the files are applicable. The data files are organized as shown in the Figure 2(b) where it shows the pilot excel books and the data files within a folder.
\begin{figure}[h]
\center
\includegraphics[width=260pt,height=150pt]{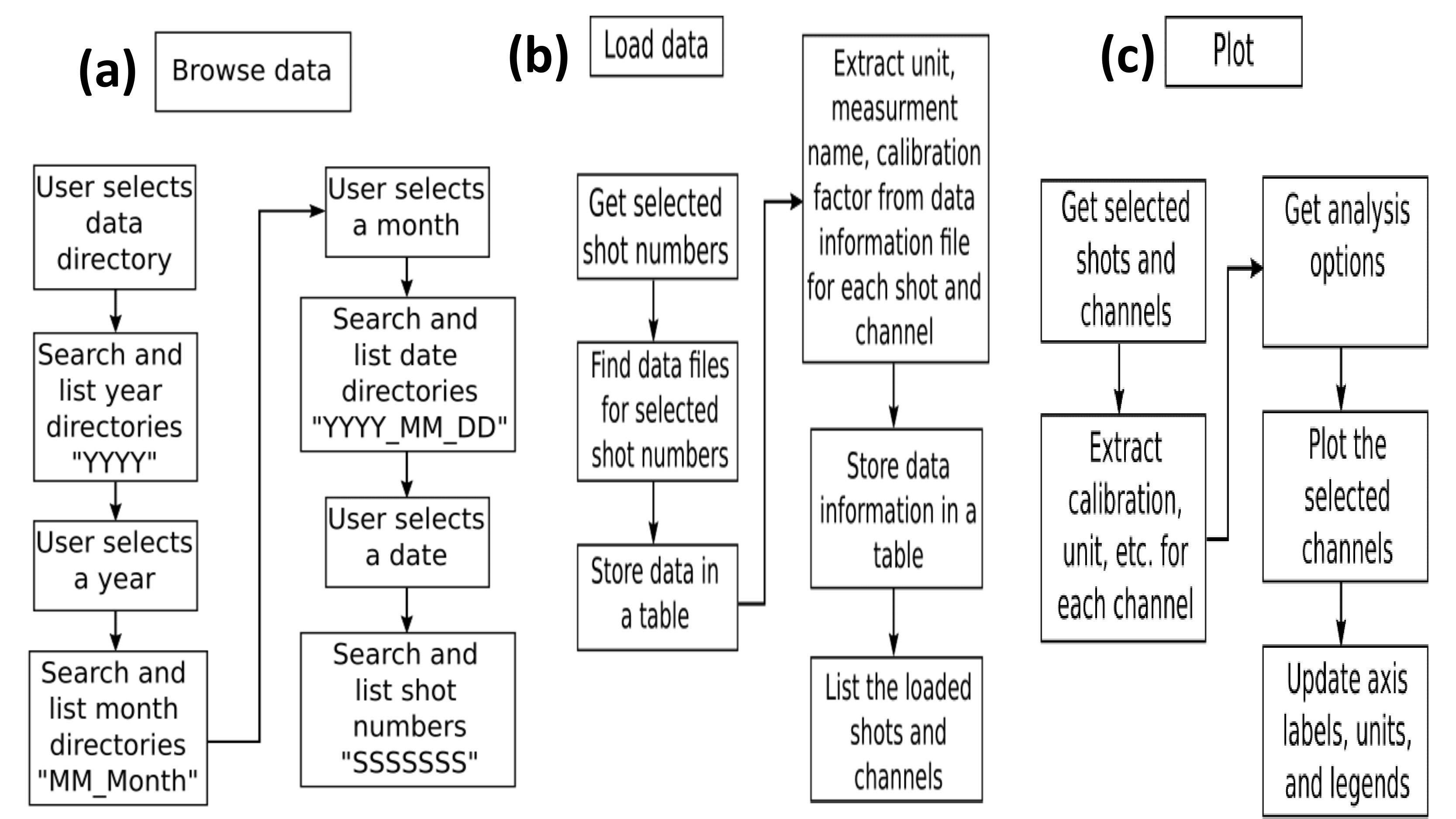}
\caption{It presents flow charts for (a)browsing data, (b)loading data, (c) generating basic plot.} \label{fig:3}
\end{figure}
\par The code starts to load experimental data in the GUI window shown in figure 1 through navigation of the proper directory where the experimental data of interests were stored. This process is done through the logical flows of `browse data', shown in figure 3(a). Figure 3(b) explains the logical flow chart of the `load data' in GUI window through interaction with pilot Excel book which is one of the key process of the code. Once the shot numbers are available, the shots of interest need to be "loaded" to the program through GUI window. The program finds all the data files associated with the selected shot numbers. All the files of the selected shots number of interests will be opened and stored in a table, particularly in the object \texttt{pandas.DataFrame} which is called the data table, where each time series data is given a unique data ID. After that the program opens the appropriate
Excel book and extracts information about the data such as unit, measurement name,
and calibration factor for each data time series. The parameters related to each data ID
are stored in another table which is called the parameter table. The program completes the loading
process by listing the shot numbers and measurement names or channels. After proper loading of the data files under analysis, figure 3(c) shows the work flow for the basic plots and visualization and the nature of experimental data.
\par It is very important to analyze the fundamental signals of various channels to check its proper nature.
 The program retrieve the information of relevant data and
parameters from the tables of data and parameter. The optional analysis, such
as smoothing factor, are also possible through graphical interface.
An interactive plot is then generated using
a plot item in \texttt{pyqtgraph} package. Based on the parameters, the program updates the
unit, axis labels and the legends.
\par Here, the most important feature of the code is successful using the in built ROI tool of \texttt{pyqtgraph} which allows to indicate region of interest (ROI) and corresponding analysis of the focussed region of data.  The ROI can be changed in an interactive fashion where it can be changed either by modifying the width of ROI or moving the ROI without changing its width. The corresponding new plot is then updated automatically. This mapping is done through a function to a signal emitted by the ROI object. In this particular case, the connected function simply plots the
data within the selected region. In our application, the ROI tool is a very important feature to
carry out spectral analysis which allows quick and detailed analysis of modes
present in the signals. The ROI tool have been used to efficiently identify GAM mode within a time series.
\par To explain GAM features, Power spectral density (PSD) and Cross-power Spectral analysis such as Cross Spectral Density(CSD), Cross Coherence and Cross Phase are essential analyzing parameters which are adopted in the code through GUI taking from Python library. Figure 4(a) and 4(b) present the logic flow chart of the PSD and Cross spectral analysis.

\begin{figure}[h]
\center
\includegraphics[width=260pt,height=150pt]{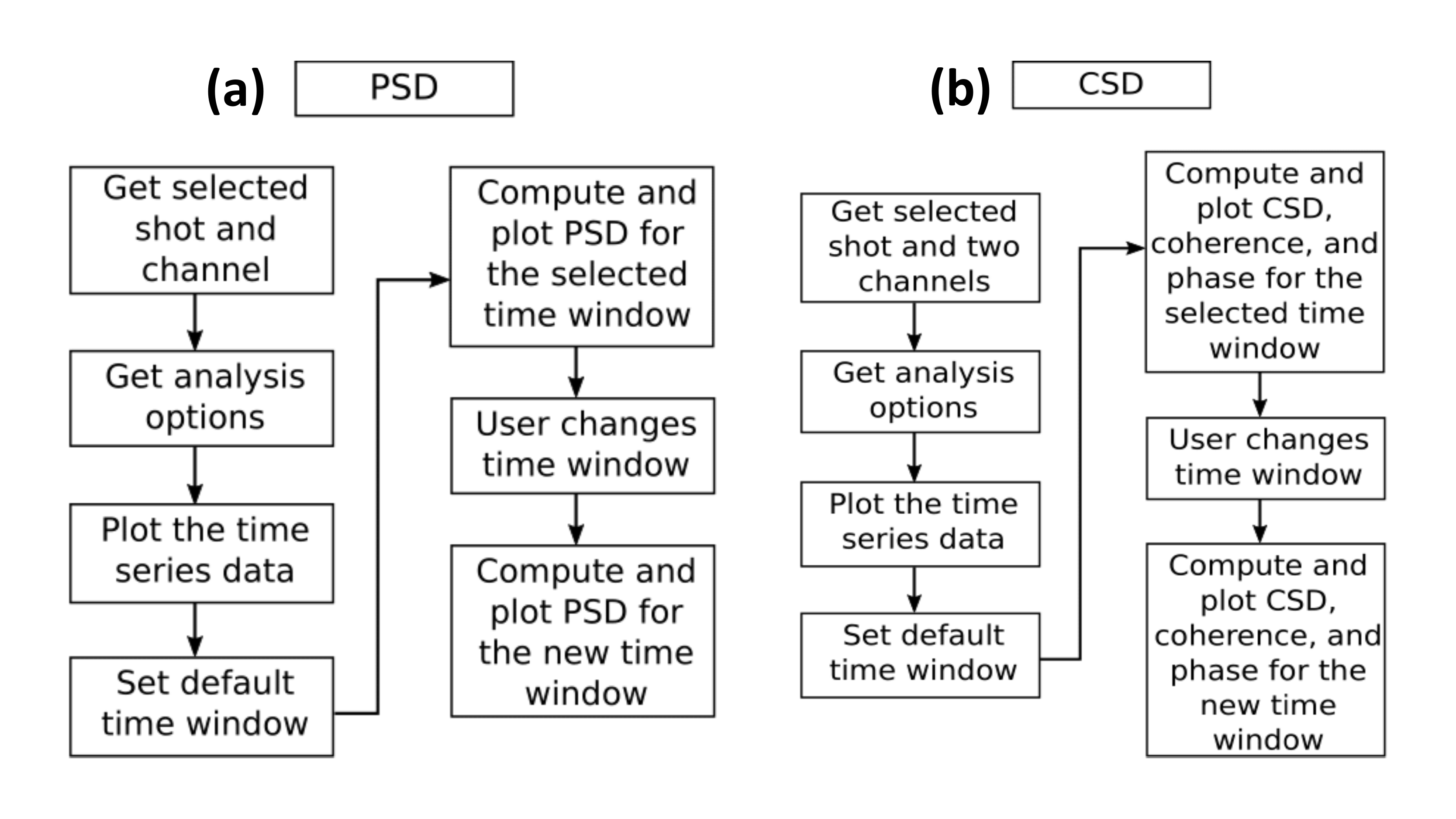}
\caption{Flow charts of (a)Power Spectral Density(PSD) and (b)Cross Spectral Analysis.} \label{fig:4}
\end{figure}
\par The analysis options for PSD such as segment length and overlap percentage are
read from the graphical interface.  A default time window is set and visualized
as a colored ROI object on the plot of the time series. The PSD is computed using a function
in \texttt{scipy} package and plotted in a separate window within a plot.
The time window, an object in \texttt{pyqtgraph} package, can be changed in an interactive
manner. Once the time window is changed, the PSD for the new
time range is computed and updated for the corresponding plot.
\par Cross spectral analysis is very useful for identifying common modes in two time series signals in GAM detection experiment. For this process, two time series signals, collected from two different independent channels are compared. The program also reads the analysis options from the graphical interface like `PSD' plot. Initially, the two time series
data are plotted. A default time window is set and displayed as a segment item on the
time series data, similar process for `PSD' plot. The cross spectral density, cross-coherence, and cross-phase are computed
for the time window and plotted in three separate windows within the plot.
When the time window is changed, the above three quantities are computed again for the new
time range and the corresponding plots are updated.
\section{Results and discussions:}
The previous section discussed about the how code works. Figure 5 represents one of the typical example about the basic plots and the corresponding plots of `PSD', `CSD', `Cross-coherence', Cross-phase' at ROI of the basic data plots. Figure 5(a) presents the `PSD' of the data regime with ROI of the basic plot of ion saturation current of Langmuir probe, as shown in figure 5(b). Similarly, figure 5(c), 5(d) and 5(e) present the `CSD', `Cross-coherence', Cross-phase' at ROI data regime of the two different ion saturation current signals collected from two different Langmuire Probes. Here, the ROI is marked by the gray rectangular window on the basic plots of figure 5. So, it is clear from the typical example of figure 5 that the mode can be easily determine in a time series by changing the ROI positions on the basic data plots and it is very powerful tools for GAM detections.
\begin{figure}[h]
\center
\includegraphics[width=260pt,height=150pt]{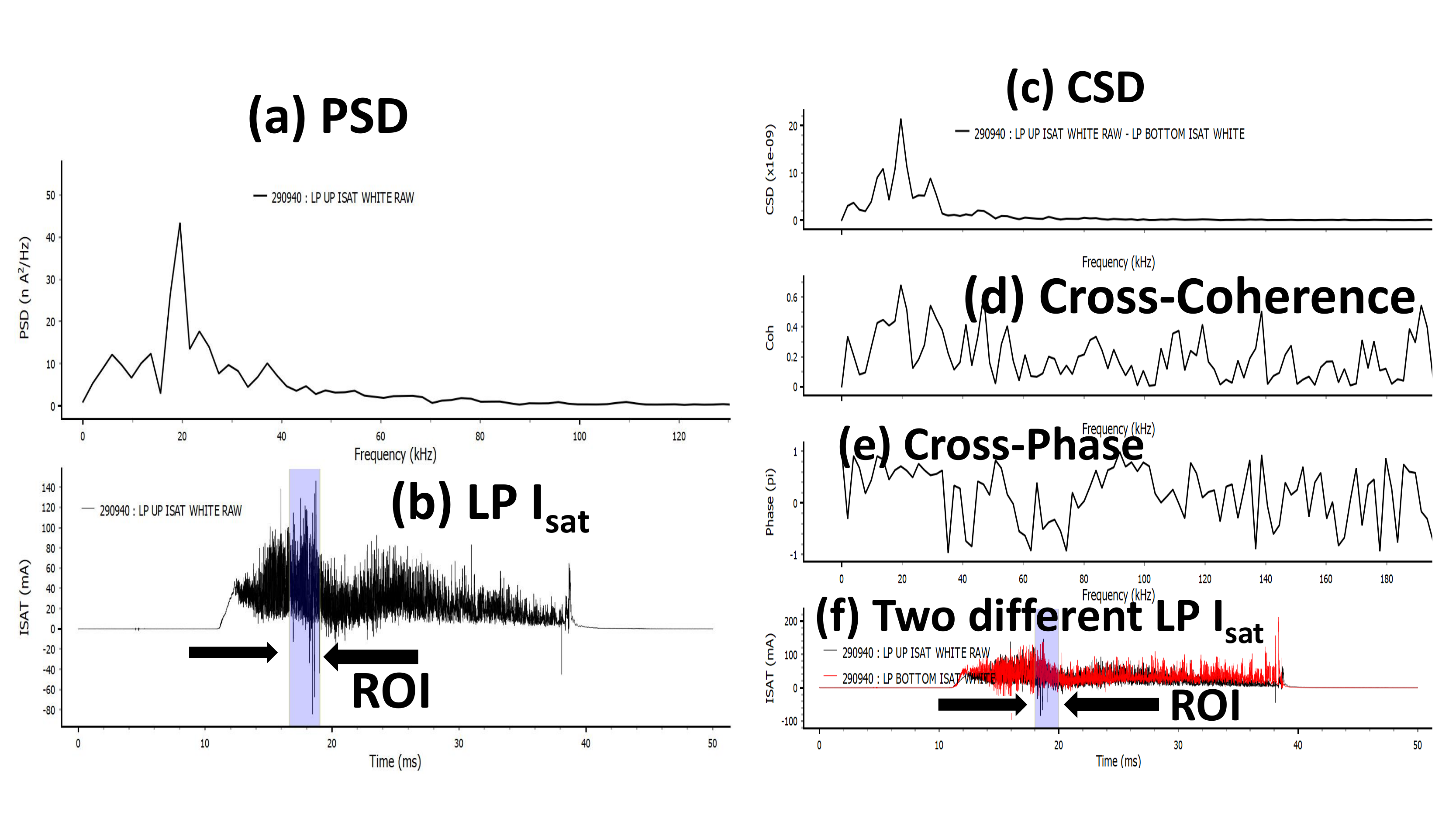}
\caption{It presents (a)Power Spectral Density(PSD) of ROI of $LP-I_{Sat}$ signal, (b)corresponding fundamental $LP- I_{sat}$ signal, (c)CSD, (d) cross-coherence, (e) cross-phase of ROI of two different $LP-I_{Sat}$ signals, (f)corresponding two $LP-I_{sat}$ signals.} \label{fig:5}
\end{figure}

\par It is well known that GAM detection experiment is a very sophisticated due to its complexity and on the other hand it is very important to understand inherent turbulent physics as well as L-H transition in Tokamaks or Stellarators. It is basically a high frequency zonal flows \cite{kn:Diamond,kn:Diamond,kn:Fujisawa,kn:DB,kn:Melnikov1} and normally confine within a small radial width. Also, recently it is found that it can be a global eigenmode as well which exhibits long-range spatial correlation\cite{kn:Melnikov13} but not in all Tokamaks or Stellarators. So, it is necessary to detect experimentally the GAM mode collecting the information from approximately same magnetic field surfaces. The same or nearly same magnitude of the $I_{sat}$ signals profiles, collected from top and bottom Langmuir Probes during the same plasma discharge serve as a criterion to indicate that all measured density fluctuations or floating potentials fluctuations are accumulated from same or nearly same magnetic field surface.

\begin{figure}[h]
\center
\includegraphics[width=260pt,height=150pt]{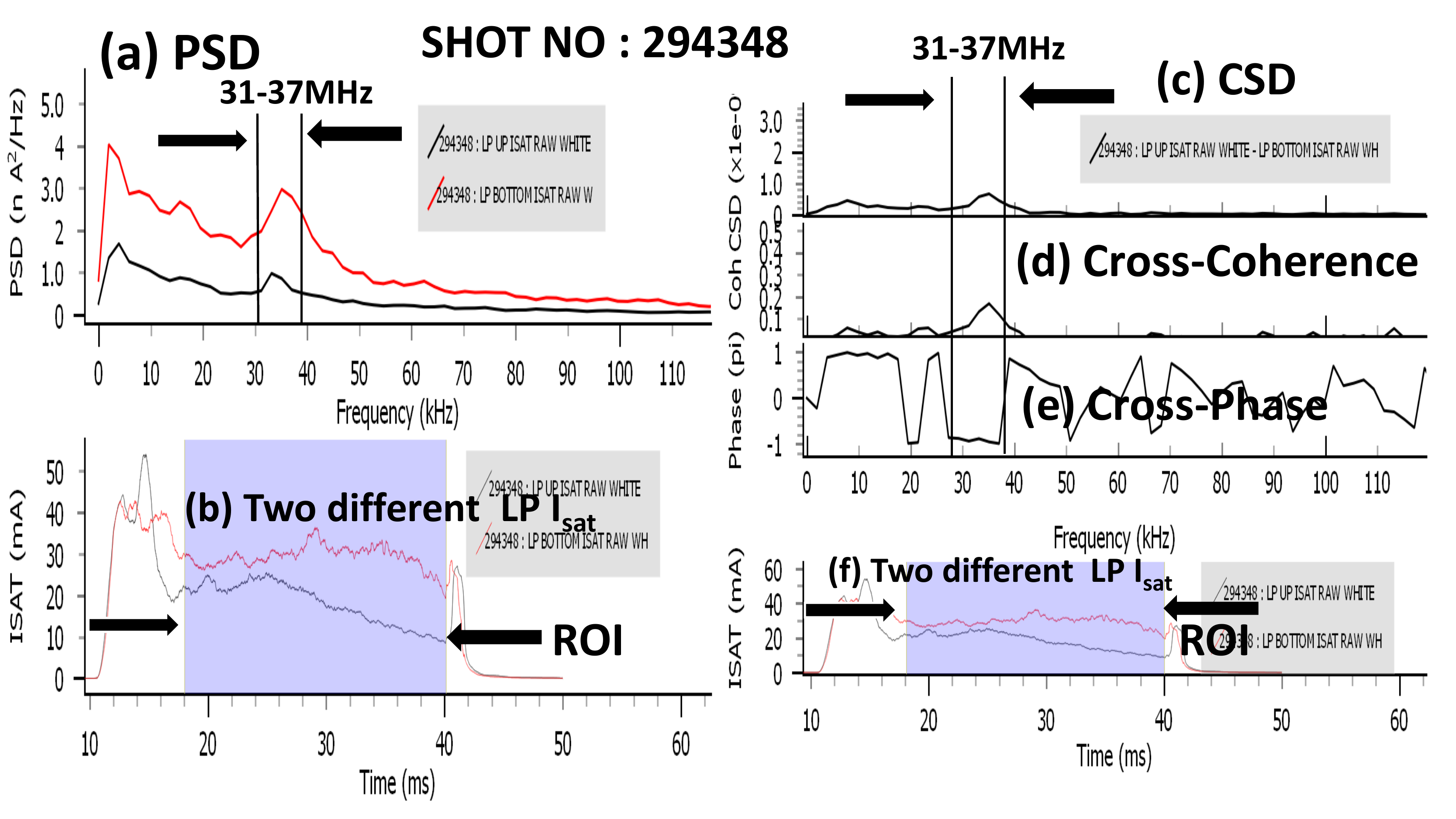}
\caption{An indication of GAM mode with frequency width 31-37MHz appears within ROI having time window 22ms as indicated by (a)Power Spectral Density(PSD) of ROI of $LP-I_{Sat}$ signal, (b)corresponding two $LP- I_{Sat}$ signals, (c)CSD, (d) cross-coherence, (e) cross-phase of ROI of two different $LP-I_{Sat}$ signals, (f) corresponding two $LP-I_{Sat}$ signals.} \label{fig:6}
\end{figure}

\par At first, ROI has been taken over whole the flat top region of $I_{sat}$ signals, collected by top and bottom Langmuir probes of a typical STOR-M plasma discharge with shot number 294348 to get an overall scenario. Figure 6 clearly indicates the presence of GAM mode at flat top where the ROI time window is 22 ms. It shows the GAM-like mode having frequency width 30-40MHz in the plasma discharge which is observed from `PSD', `CSD', `Cross-coherence' and `Cross-phase' which are presented in figure 6(a), 6(c), 6(d), 6(e) respectively. Figure 6(b) and figure 6(f) presents the similar $I_{sat}$ signals collected from `top' and `bottom' Langmuir Probes.

\begin{figure}[h]
\center
\includegraphics[width=260pt,height=150pt]{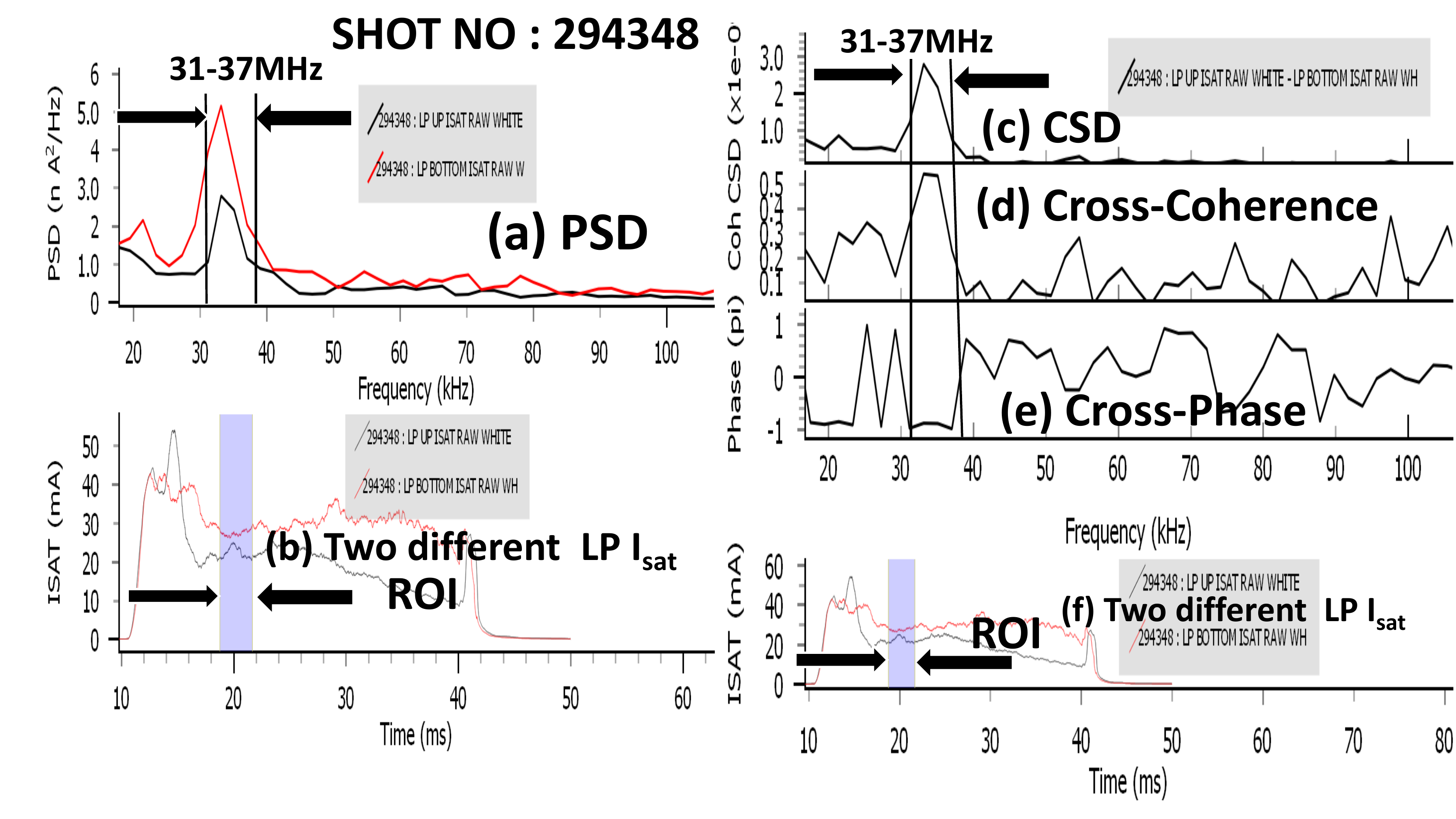}
\caption{A typical GAM mode with frequency width 31-37MHz appears within ROI as indicated by (a)Power Spectral Density(PSD) of ROI of $LP-I_{Sat}$ signal, (b)corresponding two $LP- I_{Sat}$ signals, (c)CSD, (d) cross-coherence, (e) cross-phase of ROI of two different $LP-I_{Sat}$ signals, (f) corresponding two $LP-I_{Sat}$ signals.} \label{fig:7}
\end{figure}

\begin{figure}[h]
\center
\includegraphics[width=260pt,height=150pt]{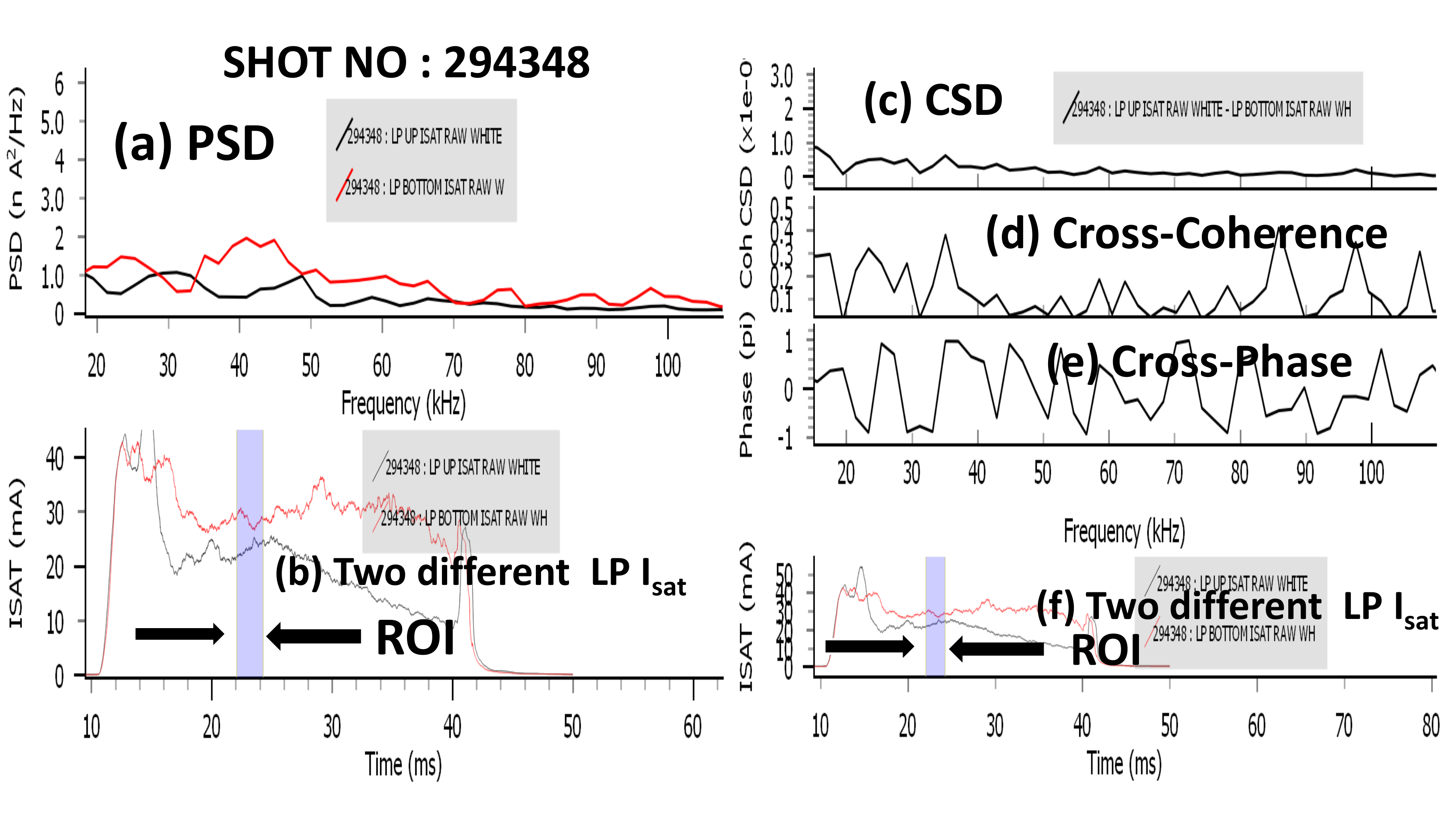}
\caption{GAM mode with frequency width 31-37MHz disappears at another ROI indicated by (a)Power Spectral Density(PSD) of ROI of $LP-I_{Sat}$ signal, (b)corresponding $LP- I_{Sat}$ signal, (c)CSD, (d) cross-coherence, (e) cross-phase of ROI between two different $LP-I_{Sat}$ signals, (f) corresponding two $LP-I_{Sat}$ signals.} \label{fig:8}
\end{figure}

\par  The ROI window has been set and started to move in different parts of those two top and bottom Langmuir probes $I_{sat}$ signals to characterize and to understand more precisely at which time GAM mode appears as well as its time duration. Figure 7 indicates that a clear GAM mode of frequency width 31-37MHz emerges around 18.8 ms within a time window of of width at around 2.9 ms which is clearly marked by a gray rectangular area of ROI. Here, figure 7(b) and figure 7(f) show similar $I_{sat}$ signals collected from `top' and `bottom' Langmuir probes which have been smoothed to get clear envelope of those signals. The evolution of the envelopes of signals suggests that they are located approximately on the same magnetic surfaces. The sharp peaks of `PSD', `CSD', `Cross-coherence' and `Cross-phase', presented in figure 7(a), 7(c), 7(d), 7(e) respectively, show clear indication of GAM-like mode when the time range of ROI at around 18.8ms within a time window of at around 2.9 ms.
\par When the ROI window is moved in different locations for the same two signals, it has been noticed that the GAM mode disappeared as shown in figure 8 for ROI window within 22-24ms on the $I_{sat}$ signals, due to the lack of the clear peaks of GAM-like mode in the `PSD', `CSD', `Cross-coherence' and `Cross-phase', presented by the figure 8(a), 8(c), 8(d), 8(e) respectively.
\par So, it is cleared that the moving of ROI on the signals can easily detect the GAM mode very precisely in terms of its time of appearance in the discharge as well as its existing time duration through the quick and efficient characterizations of the GAM mode.
\section{Conclusion and future work:}
So, it is cleared from previous discussions that this code is highly efficient to analyze and detect the sophisticated GAM-like mode. This code is user friendly through its graphical interface. It has also options to add more analytical options or components on its GUI window. Recently, the Bi-coherence analysis has been added into the codes with its flexible parameters options such that it can easily calculate higher order coherence. This addition will be tested and validated soon. In future, the more analytical options related to time series signal processing will be added and will be used for more elaborate and larger scenario.
\begin{acknowledgments}
 We would like to acknowledge NSERC for supporting this work. We also would like to acknowledge the kind support and inspiration provided by Prof. Akira Hirose. The work of A.V. Melnikov was supported in part by the Competitiveness Programme of the National Research Nuclear University ``MEPhI".
\end{acknowledgments}

\end{document}